\documentclass[sigconf]{acmart}

\usepackage{ragged2e}
\usepackage{array}
\usepackage{booktabs}

\AtBeginDocument{%
  }

\copyrightyear{2025}
\acmYear{2025}
\setcopyright{acmlicensed}\acmConference[UbiComp Companion '25]{Companion of the 2025 ACM International Joint Conference on Pervasive and Ubiquitous Computing}{October 12--16, 2025}{Espoo, Finland}
\acmBooktitle{Companion of the 2025 ACM International Joint Conference on Pervasive and Ubiquitous Computing (UbiComp Companion '25), October 12--16, 2025, Espoo, Finland}
\acmDOI{10.1145/3714394.3756342}
\acmISBN{979-8-4007-1477-1/2025/10}

\begin{document}

\title{Context-Aware Intelligent Chatbot Framework Leveraging Mobile Sensing}

\author{Ziyan Zhang}
\email{18968175831@163.com}
\affiliation{%
  \institution{Nankai University}
  \state{Tianjin}
  \country{China}
}

\author{Nan Gao}
\email{nan.gao@nankai.edu.cn}
\affiliation{%
  \institution{Nankai University}
  \state{Tianjin}
  \country{China}
}

\author{Zhiqiang Nie}
\email{nzq24@tsinghua.edu.cn}
\affiliation{%
  \institution{Tsinghua University}
  \state{Beijing}
  \country{China}
}

\author{Shantanu Pal}
\email{shantanu.pal@deakin.edu.au}
\affiliation{%
  \institution{Deakin University}
  \state{Melbourne}
  \country{Australia}
}
\author{Haining Zhang}
\email{zhanghaining@nankai.edu.cn}
\affiliation{%
  \institution{Nankai University}
  \state{Tianjin}
  \country{China}
}
\begin{abstract}
With the rapid advancement of large language models (LLMs), intelligent conversational assistants have demonstrated remarkable capabilities across various domains. However, they still mainly rely on explicit textual input and do not know the real world behaviors of users. This paper proposes a context-sensitivetive conversational assistant framework grounded in mobile sensing data. Collecting user behavior and environmental data through smartphones, we abstract these signals into 16 contextual scenarios and translate them into natural language prompts, thus improving the model's understanding of the user’s state. We design a structured prompting system to guide the LLM in generating a more personalized and contextually relevant dialogue. This approach integrates mobile sensing with large language models, demonstrating the potential of passive behavioral data in intelligent conversation and offering a viable path toward digital health and personalized interaction.
\end{abstract}




\maketitle

\section{Introduction}
In recent years, large language models (LLMs), such as, \textit{ChatGPT} and \textit{Gemini} have reshaped human–computer interaction with their natural, fluent and context-aware conversational capabilities \cite{EatingDisorder}. They have found widespread application in domains, such as, education, healthcare, and productivity, drawing significant attention from both academia and industry. Studies have shown that LLM-based conversational assistants perform well in task assistance, emotional support, and decision-making recommendations \cite{MemoryReviver} \cite{ZzzMate}.
However, by relying solely on explicit textual input, current conversational agents lack awareness of users’ real-time behaviors and environmental contexts, thereby limiting their capacity for inferring implicit needs and providing personalized, proactive interactions. Previous research \cite{ondevicelanguagemodelscomprehensive}\cite{Yang} lacks comprehensive frameworks for contextual modeling or focuses narrowly on isolated components such as behavior recognition or language optimization, without effectively integrating real-world behavioral sensing. Consequently, current conversational agents remain significantly limited in both understanding the user context and enabling contextually nuanced interactions. This critical limitation highlights the transformative potential of mobile sensing technologies, which enable continuous and unobtrusive monitoring of real-world contexts - precisely identifying \textit{user locations} and \textit{ongoing activities} throughout daily life \cite{Harari2023}, thus offering a promising pathway toward genuinely context, aware conversational systems.

To address these challenges, we propose a context-aware conversational assistant framework that integrates mobile sensing data. The framework is based on the hypothesis that a deeper understanding of the fundamental states of users, including location, activities, and environmental contexts, can significantly enhance the relevance and effectiveness of the interaction. Using smartphone sensors (such as, GPS, screen activity detection, motion sensors, application usage monitors), the system passively collects multimodal behavioral data. To systematically organize this complex behavioral information, we construct five core behavioral categories encompassing 16 specific contextual scenarios. These abstracted scenarios are then translated into natural language prompts and incorporated into the large language model's input. This integration enables the assistant to generate responses informed not only by textual input but also by users' real-world contexts, facilitating more proactive and personalized dialogue generation.

To evaluate the feasibility of the proposed framework, we developed a prototype system and conducted a small-scale pilot study to explore the impact of behavioural signals (such as, app usage patterns, screen activity, and mobility) on conversational outcomes. We also discuss the implications of our proposed framework, which sheds light on how integrating mobile sensing with large language models can enable more natural and proactive human-computer interaction, with significant potential for digital health applications and personalized behavioral interventions.
The main contributions of this paper are as follows:

\begin{enumerate}
  \item We propose a lightweight architecture that integrates mobile sensing data with LLMs, enabling behavior-context-aware dialogue generation.
  \item We design a generic representation of sensed contexts and establish a bridge between sensor data and natural language prompts.
  \item We provide preliminary empirical evidence demonstrating the potential of this method to enhance personalization and proactivity in dialogue systems.
\end{enumerate}

We argue this approach holds a significant promise for applications in digital health, psychological support, and behavioural interventions, and contributes to the advancement of human–computer interaction systems grounded in real-world context awareness.

\section{Related Work}

This section reviews most relevant research works to our work. We first examine key characteristics and limitations of existing LLM-based dialogue systems (\S 2.1), then discuss the context-aware approaches—particularly mobile sensing techniques—for enhancing conversational Artificial Intelligence (AI) (\S 2.2).

\subsection{Forms and Applications of LLM-based Conversational Assistants}

With the rapid development of LLMs such as the GPT series,Grok, and Gemini, intelligent conversational assistants have been widely applied in domains including question answering, content generation, emotional interaction, and education, significantly enhancing the human–computer interaction experience. Commercial products such as Copilot, ChatGPT, and Bard have demonstrated the powerful capabilities of LLMs in multi-turn dialogue and language understanding and generation.
Nevertheless, these assistants typically rely on explicit user input in the form of text and lack the ability to perceive users’ current behaviors and surrounding environments \cite{gao-etal-2018-neural-approaches}. As a result, they are limited in identifying users’ implicit needs proactively, thereby constraining the depth of personalization and contextual understanding. Existing research primarily focuses on improving dialogue generation quality, knowledge integration, and multimodal processing, with relatively little attention paid to incorporating real-world user behavior data to support more natural interactions.

In recent years, multimodal foundation models, such as, GPT-4, Flamingo\cite{NEURIPS}, and PaLM-E\cite{PaLM} have also gained attention to their ability to process diverse input modalities, including images and speech, thereby expanding both the modalities and capabilities of interaction. However, these models still mainly depend on explicit user input and lack continuous perception of user behavior and context \cite{baltruvsaitis2019multimodal}. Additionally, their substantial computational requirements \cite{Transformers} and potential privacy concerns pose challenges for real-time deployment in mobile scenarios.
Therefore, how to effectively integrate real-life behavioral and contextual information into conversational systems, enabling more proactive and personalized services, remains a critical and open challenge.

\subsection{Mobile Sensing Technologies and Applications}

Modern smartphones and wearable devices are equipped with a variety of sensors, such as, GPS, accelerometers, gyroscopes, and microphones, making mobile sensing a key method for understanding users’ behaviors and states. Prior studies have demonstrated that these sensor signals can be used to infer emotions \cite{gao2023living}, activity types, social interactions, sleep patterns, and psychological stress levels, with wide applications in health monitoring and behavioral analysis.

Some recent works have explored integrating mobile sensing data with LLMs. For instance, Zhang et~al. \cite{PredictAffective} utilized the Aware-Light platform to collect raw sensor data, which was then processed and fed into a language model to perform preliminary user state modeling. However, this approach places high demands on model capacity and input length, often failing to fully extract deep semantic information from the data. It also lacks structured guidance for the model, resulting in limited interaction quality and naturalness. Gao et al. \cite{gao2024leveraging} proposed an automated mobile sensing strategy leveraging LLMs, to address the inefficiencies and lack of generalizability in traditional mobile sensing for human behaviour understanding.

Zhang et al. proposed a privacy-preserving, locally deployed framework that integrates multimodal smartphone sensors with LLMs \cite{OnDeviceLLMs}, achieving context-aware recommendations. Nevertheless, their approach exhibits limited generalizability in scenario modeling and lacks flexibility in dialogue interaction. The ScreenTK system \cite{ScreenTK} leverages screen content in combination with LLMs to detect “time-killing” behaviors, significantly improving behavior recognition accuracy, but it does not incorporate behavior context into conversational content. Additionally, Chen et al. \cite{qin2024ondevice} proposed a self-supervised data selection and synthesis framework for efficient on-device personalized fine-tuning, yet it does not involve behavior-level sensing, thus limiting its ability to achieve deeper contextual understanding.

In summary, unlike the above mentioned works, we propose an enhanced approach: sensor data collected from smartphones is interpreted through a set of generalized contextual scenarios, and transformed into natural language prompts to guide the language model in understanding the user’s state. Furthermore, we design a hierarchical prompting structure to improve the efficiency of data interpretation and the model’s capacity for context comprehension. We also expand the modes of interaction, such as incorporating emoticons and speech synthesis, to enhance user engagement and emotional expressiveness during interaction.

\begin{table*}[t]
\centering
\caption{Sensor data collection specifications.}
\label{tab:data_collection}
\setlength{\tabcolsep}{4pt}
\small
\begin{tabular}{
    >{\RaggedRight}p{3cm}
    >{\RaggedRight}p{9cm}
    >{\RaggedRight}p{2.5cm}
}
\toprule
\textbf{Data Type} & \textbf{Fields} & \textbf{Frequency} \\
\midrule
Activity Data & activity (String) - Recognized activity type (walking, running, cycling) & 500 steps \\
Accelerometer & x,y,z (Float) - Acceleration values & 50Hz \\
Battery & level (Float) - Battery percentage & 5 sec \\
Bluetooth Devices & pcCount, phone Count (Int) - Device counts & 60 sec \\
Bluetooth Status & status (String) - Connection state & Event \\
Gyroscope & x,y,z (Float) - Rotation rates & 50Hz \\
Light & light Level (Float) - Light intensity (lux) & 5Hz \\
Location & lat,long (Double) - Coordinates; location Type (String) & 5 sec \\
Screen Status & screen Status (String) - Screen state & Event \\
Wifi Status & status (String) - Connection state & Event \\
Audio & audio Device (String) - Output type; isActive (Boolean) & 5 sec \\
App Usage & appName, package Name (String); duration (Long) & 30 sec \\
Foreground App & package Name, appName (String) - Foreground app & 30 sec \\
\bottomrule
\end{tabular}
\end{table*}

\section{Method}

This section presents the technical foundations of our context-aware chatbot framework. We begin by detailing the sensor infrastructure for multimodal data acquisition, followed by our approach to abstracting sensor signals into contextual scenarios. We then describe dialogue optimization techniques for enhanced user interaction, and conclude with the system implementation details. 

\subsection{Sensor Design}

 The system employs multiple types of sensors to achieve comprehensive perception of user behaviors and environmental conditions, integrating three categories of sensors: hardware sensors, software sensors, and context sensors. These sensors collectively capture both user activities and situational states \cite{HumanActivity}.

\begin{itemize}
  \item \textit{Hardware Sensors:} This category includes linear accelerometers and gyroscopes for recognizing activities such as running and walking, temperature sensors for detecting ambient temperature, and light sensors for distinguishing indoor versus outdoor environments.
  \item \textit{Software Sensors:} These monitor system-level information such as application usage, battery status, audio playback, and network connectivity.
  \item \textit{Context Sensors:} This includes location data (such as, GPS), screen status (such as, screen on/off or unlocked), and peripheral device awareness (such as, Bluetooth/Wi-Fi).
\end{itemize}

Data collection is performed through event listeners or periodic polling mechanisms, with the gathered data stored locally in a database to support subsequent behavior modeling and contextual event triggering. The specifications of collected data types and their corresponding collection frequencies are presented in Table~\ref{tab:data_collection}.

\subsection{Contextual Scenario Design}

Based on user interviews and behavioral research, the system defines sixteen common contextual scenarios designed to accurately capture user behavioral states\footnote{This study involved 19 participants and conducted interviews on multiple core issues. The collected data serves multiple research purposes, while this study specifically focuses on a subset of the data, with other portions not elaborated here.}. These scenarios serve as triggers for corresponding interactive prompts. Scenario triggering relies on multimodal sensor data fusion, model prediction outputs, and user behavior modeling. A broadcast mechanism informs the main application when a scenario is activated, and the main application sends corresponding trigger phrases to the LLM, which then generates personalized responses for display to the user. Classification and examples of these scenarios are presented in Section 4.

\subsection{Dialogue Optimization}
Dialogue optimization plays a crucial role in enhancing user experience and strengthening the interaction between the application and its users. This section focuses on four key areas: sentence segmentation, speech synthesis, the use of stickers (emoticons), and emoji feedback. These functions are systematically optimized and deeply integrated from the perspectives of enhancing voice interaction, enriching emotional expression, and increasing user engagement.

\begin{itemize}
    \item \textit{Sentence Segmentation}: Dividing lengthy responses into sequential phrases for improved readability
    \item \textit{Speech Synthesis}: Implementing neural text-to-speech with buffering for natural voice output
    \item \textit{Sticker Integration}: Generating contextual emoticons through sentiment analysis
    \item \textit{Emoji Feedback}: Enabling quick reaction logging via interaction gestures
\end{itemize}

First, the sentence segmentation technique breaks down long sentences into structurally clear short sentences, enabling the LLM to process information sequentially. This approach prevents information overload and semantic ambiguity, while improving the accuracy of speech-to-text transcription and the fluency of speech playback. Such segmentation not only simulates the natural rhythm of human conversations but also reduces the cognitive load on users.
Second, speech synthesis employs Microsoft Azure’s multilingual neural network technology, combined with voice prewarming and asynchronous processing strategies. This ensures that voice output is natural, smooth, and responsive, without compromising the fluidity of text input and output.
Third, the sticker function automatically generates keywords through sentiment analysis and invokes sticker APIs to add real-time emotional nuance to the conversation. This enhances emotional conveyance and user interaction experience, especially when the LLM’s response is not segmented, as stickers effectively enrich the dialogue content.
Finally, the Emoji feedback feature implements an interaction flow of “long press message → display emojis → select feedback”, allowing users to quickly express their emotional attitude toward the bot’s response. These feedback data are stored for subsequent statistical analysis, which not only increases user engagement but also provides valuable insights for the continuous improvement of human-computer dialogue quality.

Overall, these four optimization methods work synergistically to significantly enhance the system’s natural interaction capability and user satisfaction.

\begin{figure}[t]
  \centering
  \includegraphics[scale=0.26]{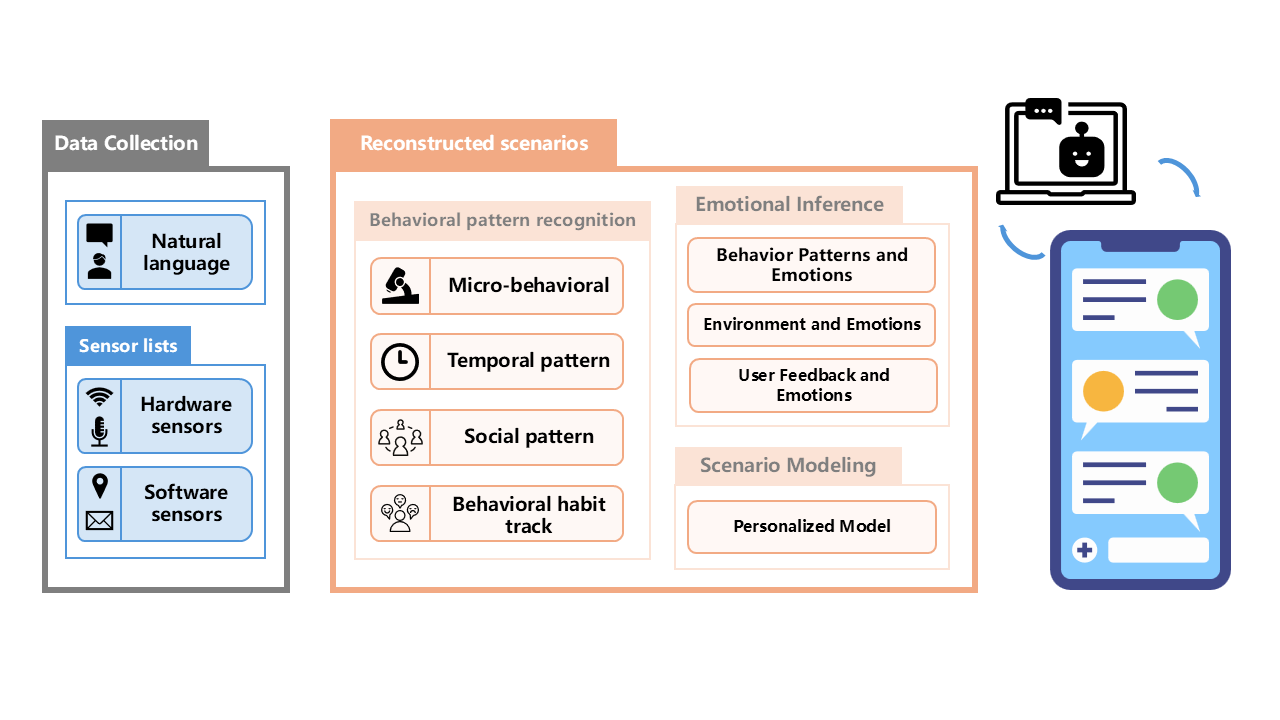}
  \caption{The  system architecture of our context-aware chatbot framework. The architecture demonstrates the complete data flow from mobile sensing collection through contextual scenario abstraction to personalized dialogue generation.}
  \label{fig:architecture}
\end{figure}

\subsection{Technical Implementation }
This section discusses the technical inplementation. The mobile application was developed for the Android platform using Kotlin and the Android Jetpack framework. 
Hardware sensors including the accelerometer, gyroscope, and light sensor were initialized via the \texttt{SensorManager} method, with sampling rates set at 50Hz for motion sensors and 5Hz for the light sensor. System-level data collection leverages Android framework services: application usage statistics are polled every 30 seconds using \texttt{UsageStatsManager}, battery status is monitored at 5-second intervals through \texttt{BatteryManager}, audio playback state is tracked via \texttt{AudioManager}, location updates occur every 5 seconds using fused location providers, screen state transitions are captured through broadcast receivers, and network connectivity is periodically scanned via \texttt{ConnectivityManager}. All raw sensor data undergoes preprocessing for noise reduction before storage.

The Room persistence library provides structured local storage with efficient query capabilities. The synchronization mechanism employs a periodic incremental strategy: daily tasks scheduled by \texttt{WorkManager} identify new records since the last synchronization timestamp, serialize data into JSON format, and transmit via Retrofit with JWT authentication in the \texttt{Authorization: Bearer} header. The Flask-based backend implements a multi-tenant architecture where each user maintains an isolated database for sensor and interaction data, while the main database stores authentication credentials.
Building upon this robust sensing infrastructure, our system (as illustrated in Fig.~\ref{fig:architecture}) integrates the aforementioned multimodal sensor data with user interaction patterns. Specifically, the framework leverages generic behavioral scenarios to recognize user activities, then employs contextual analysis to infer emotional states based on situational cues. This integrated processing pipeline enables deep, contextually-aware interactions between the LLM and end-users through personalized conversational engagements.

We integrated our structured prompts with the ChatGPT-4o model, a multimodal large language model developed by OpenAI. Its fast response time and strong contextual reasoning capabilities make it suitable for real-time interaction tasks in mobile environments.

\section{Scenario Design} 

Based on user interviews and behavioral studies, this section designs 16 common scenarios aimed at accurately capturing user states. These scenarios leverage multi-sensor data fusion, model prediction, and behavioral modeling to trigger corresponding interactions. The system broadcasts trigger keywords to the LLM, which generates personalized responses and presents them to the user, thereby enhancing the intelligence and relevance of the interaction.

\subsection{Scenario Categories}

This section categorizes 16 contextual scenarios into four functional domains based on behavioral traits and intervention goals. The classification framework organizes scenarios based on shared behavioral characteristics and intervention objectives:

\begin{itemize}
  \item \textit{Exercise/Activity:} Includes walking detection, running detection, intense exercise detection, prolonged sitting reminders, etc.
  \item \textit{Time/Routine:} Includes nap detection, wake-up detection, insomnia detection, meal pattern monitoring, nighttime automatic analysis, etc.
  \item \textit{Location/Environment:} Includes workplace detection, off-work detection, travel recommendation detection, etc.
  \item \textit{Usage Behavior/Media:} Includes excessive app usage detection, music playback detection, story reminders, late-night binge-watching detection, etc.
\end{itemize}

\subsection{Representative Scenarios}

To illustrate how our system translates raw sensor data into meaningful contextual interactions, we detail one representative and high-impact scenario: \textit{Excessive App Usage Detection}. This example demonstrates how multimodal sensing, behavioral modeling, and structured prompt engineering are integrated to enable empathetic, context-aware dialogue.

This scenario targets late-night overuse of attention-draining applications such as TikTok or Weibo, which may be linked to negative emotional states or disrupted sleep routines. The system continuously monitors multiple sensor streams, including:

\begin{itemize}
  \item \textit{App usage logs:} duration and frequency of specific app use.
  \item \textit{Timestamps:} especially between 00:00 and 06:00.
  \item \textit{Location data:} to determine whether the user is stationary or at home.
  \item \textit{Screen status:} to track whether the screen is actively on for prolonged periods.
\end{itemize}

When the user exhibits a pattern of prolonged social media use, such as, exceeding two hours of cumulative usage during late-night hours, the system triggers the \textit{“excessive usage” context}. This context activates a structured prompt that is sent to the LLM. The prompt includes the following components:

\begin{itemize}
  \item \textit{Role:} ``You are a behavior-aware companion who understands user habits. You provide gentle emotional support and insight when users show signs of app overuse, without being intrusive.''
  \item \textit{Task:} ``Reflect with the user on their recent phone usage. Suggest small, empathetic nudges such as breathing exercises, a break, or sleep preparation tips.''
  \item \textit{Requirement:} ``Be non-judgmental, supportive, and emotionally intelligent. Encourage reflection rather than enforcement.''
  \item \textit{Style Reference:} ``Friendly, calm, and conversational. Avoid alarmist tones.''
\end{itemize}

The LLM then combines these prompt instructions with recent user interaction history and real-time usage signals to generate a contextually sensitive message. For example, it may say:

\begin{quote}
\emph{``It seems like you've been scrolling for quite a while—maybe your mind is trying to unwind? Just checking in—how are you feeling right now? If you're tired but restless, I can suggest something relaxing''.}
\end{quote}

This message is further enriched through features like emoji feedback and optional speech synthesis, encouraging natural user engagement. By delivering reflective, non-intrusive prompts that resonate with the user's current behavior and emotional state, the system fosters a sense of companionship while promoting healthier digital habits.

\begin{figure}[t]
  \centering
  \includegraphics[scale=0.26]{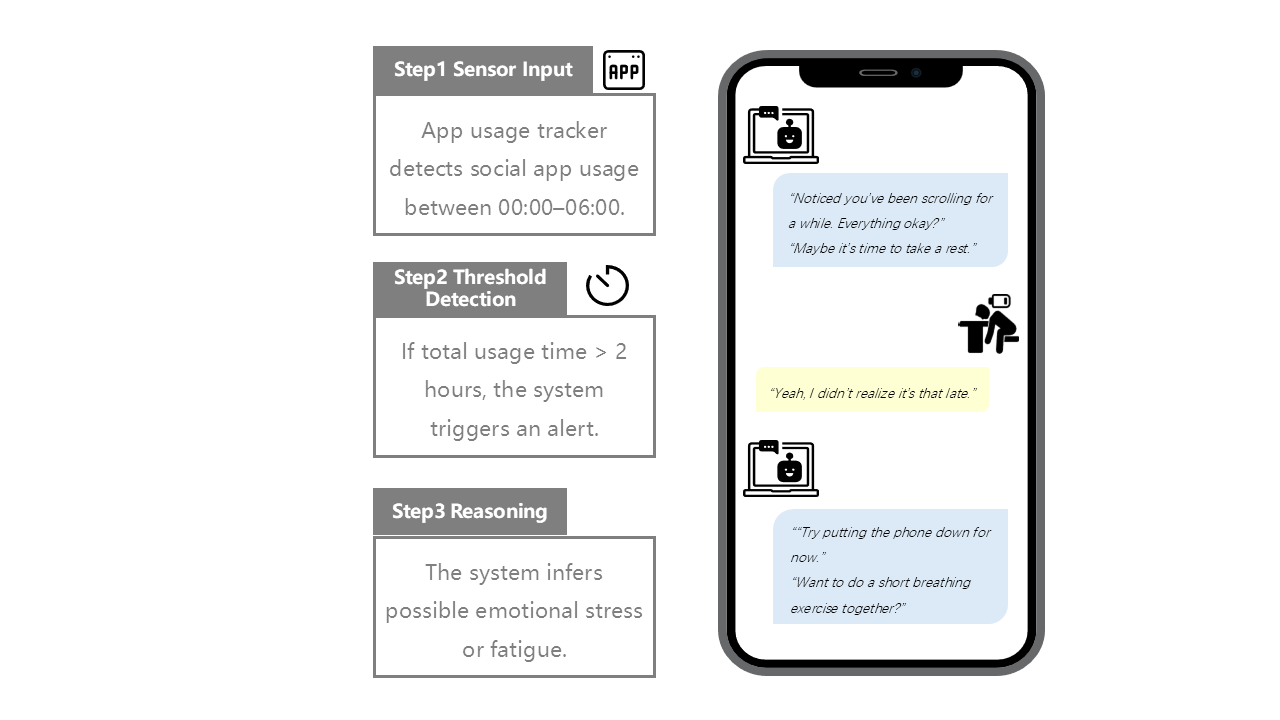}
  \caption{This figure presents an example of how the excessive app usage detection scenario operates in practice. The figure demonstrates the system's ability to monitor late-night social media usage patterns and generate contextually appropriate interventions. When the system detects prolonged engagement with social applications during nighttime hours (midnight to 6 a.m.), it triggers a gentle, caring reminder that acknowledges the user's current state while encouraging healthier behavior patterns.}
  \label{fig:Scenario Example}
\end{figure}

\section{Results and Discussion}
\label{sec:results}

To validate the feasibility and user experience of the proposed context-aware chatbot framework, we conducted a preliminary pilot study involving five university students (three female, two male) over a two-week deployment period. Participants installed the prototype system on their Android smartphones, enabling continuous collection of sensor data (GPS, accelerometer, app usage, screen status) and interaction with the chatbot. Below, we summarize the key observations and user feedback.

\subsection{Key Findings}
\label{subsec:key_findings}

Our preliminary study evealed several noteworthy insights into the effectiveness and real-world viability of the proposed system. Here, we summarize the three key findings:

\paragraph{1) Context Recognition: Mostly Reliable, With Room for Refinement.}  
The system demonstrated a solid ability to recognize user context and appropriately trigger corresponding interactions. On average, \textbf{84\%} of predefined scenarios were accurately identified across all participants. Notably, scenarios tied to screen usage and location-based events performed particularly well—\textit{Excessive App Usage Detection} reached an accuracy of \textbf{92\%}, while \textit{School/Workday End Detection} achieved \textbf{87\%}. These results highlight the robustness of using passive sensing data to infer structured, routine-based behaviors.
However, not all contexts proved equally easy to detect. For instance, \textit{Prolonged Sitting Reminders} yielded a lower accuracy of \textbf{76\%}, likely due to the inherent ambiguity in distinguishing focused sedentary activities (such as, studying) from prolonged inactivity. This finding suggests future improvements may require richer sensor fusion or personal baseline adaptation to better capture nuanced physical states.

\paragraph{2) User Engagement: Natural, Multimodal, and Appreciated.}  
Usage patterns showed that participants actively engaged with the chatbot on a daily basis, initiating an average of approximately \textbf{3.2 proactive interactions per day}. Importantly, interaction quality was enhanced by multimodal features: emoji or sticker feedback was used in \textbf{68\%} of user responses, and sentence segmentation, a feature improving readability and speech output, was employed in \textbf{41\%} of cases. These numbers suggest that users found the system approachable, expressive, and easy to use, with the added emotional cues fostering a more human-like conversational experience.

\paragraph{3) Behavioral Impact: Small Nudges, Real Effects.}  
Beyond engagement, the system also demonstrated its potential to shape user behavior. Four out of five participants reported reducing their late-night usage of social apps by \textbf{30--50\%} after receiving targeted reminders, suggesting that even gentle prompts can trigger reflection and behavioral adjustment. All participants also reported heightened awareness of their physical routines, attributing it to the system's timely and relevant feedback. One participant noted, \textit{``Seeing the alert made me realize I was scrolling mindlessly''}, reflecting a moment of personal insight that the system helped facilitate.

In summary, these findings support our core hypothesis: context-aware prompting, driven by passive sensing, can significantly improve the relevance, usefulness, and empathy of human-AI interactions in everyday settings.

\subsection{User Feedback}
Qualitative interviews revealed consistent and insightful themes across participants:
\begin{itemize}
    \item \textbf{Personalization Value}: Users appreciated that the chatbot was not offering one-size-fits-all responses, but rather adjusting naturally to what they were doing at the time:
    \begin{quote}
        ``Unlike generic chatbots, this app noticed I was running and suggested a stretching routine without me asking. It felt like a friend who pays attention.'' \\
        \hfill \textit{(Participant M2)}
    \end{quote}
    \begin{quote}
        ``The way it remembers my preferences between sessions makes interactions feel natural and personalized.'' \\
        \hfill \textit{(Participant F1)}
    \end{quote}
    
    \item \textbf{Proactive Assistance}: Several participants remarked on how the system seemed to know when to step in, offering support or check-ins at just the right moment:
    \begin{quote}
        ``The `Off-work Detection' message asking about my day came exactly when I left the lab. It was eerily timely and actually comforting.'' \\
        \hfill \textit{(Participant F3)}
    \end{quote}
    \begin{quote}
        ``Getting study break reminders just when my focus starts to fade shows it truly understands my patterns.'' \\
        \hfill \textit{(Participant M1)}
    \end{quote}
    
    \item \textbf{Usability Concerns}: A few participants pointed out moments when the system did not quite get it right, but still offered constructive suggestions for improvement:
    \begin{quote}
        ``Adding a `snooze' option for reminders would help -- sometimes I need 10 more minutes on social media.'' \\
        \hfill \textit{(Participant F2)}
    \end{quote}
    \begin{quote}
        ``When it mistook my bus ride for a running session, the fitness tips felt out of place. Location accuracy could improve.'' \\
        \hfill \textit{(Participant F3)}
    \end{quote}
\end{itemize}

\textbf{Overall Impression}: All participants rated the system \textbf{$\geq$4.2/5} for ``contextual relevance'' and ``companionship''. What stood out most in unsolicited praise was the way the chatbot seemed to fit into their lives without getting in the way:

\begin{quote}
    ``This app works because it understands my life without constant input. The reminders never felt intrusive, just smart.'' \\
    \hfill \textit{(Participant M1)}
\end{quote}

\begin{quote}
    ``It strikes the perfect balance between being helpful and not annoying. Wish more apps were like this.'' \\
    \hfill \textit{(Participant F2)}
\end{quote}

\subsection{Discussion and Limitations}
\label{subsec:limitations}

While initial user feedback was largely positive, a few  limitations emerged that warrant further attention. (i) \textit{Privacy Concerns:} Although participants consented to location tracking, three expressed discomfort with constant monitoring. This tension underscores the need for privacy-preserving designs. Future versions will prioritize on-device processing and explore federated learning to minimize cloud reliance.
(ii) \textit{Limited Participant Diversity:} The current study involved only five university students, limiting generalizability. To better assess broader impact, we plan a larger longitudinal study with 50+ participants from varied backgrounds.
(iii) \textit{Incomplete Scenario Coverage:}  
Missed interaction opportunities in unstructured contexts (such as, social gatherings or irregular travel) suggest that the 16 predefined scenarios may be insufficient. Expanding the scenario set and allowing user-defined events could enhance adaptability.
Despite these limitations, early results show that integrating mobile sensing with LLMs improves contextual sensitivity and fosters a more human-like, timely, and personalized user experience.

\section{Future Research Directions}

We identify three key research trajectories: (i) Proactive interaction systems that leverage multimodal sensing for health interventions and emotional companionship, particularly in mental health applications; (ii) Privacy-adaptive intelligence through edge-centric computation, localized model training with parameter isolation, federated learning, and differential privacy to address data sovereignty concerns while maintaining context awareness; and (iii) Ecosystem-driven platformization with standardized context-aware APIs and modular dialogue plugins for cross-domain applications in education, eldercare, and fitness.

\section{Conclusion}

This study successfully demonstrated a context-aware conversational assistant that integrates mobile sensing with LLMs, enabling proactive and personalized interactions through 16 behavioral scenarios. The system achieved 84\% context recognition accuracy and positive user feedback, particularly in digital health and attention management applications. While challenges remain in privacy protection and personalized modeling, our approach shows promising potential for driving conversational systems toward more natural, human-centered interactions. Future work will focus on edge computing, federated learning, and larger-scale user studies to enhance system adaptability and practical deployment.

\begin{acks}
This work is supported by the Natural Science Foundation of China (Grant No. 62302252).
\end{acks}

\bibliographystyle{ACM-Reference-Format}
\bibliography{ref}

\end{document}